\documentclass[10pt,conference]{IEEEtran}
\usepackage{amsmath}
\usepackage{amssymb}
\usepackage{mathtools}
\usepackage{graphicx}
\graphicspath{{./../figures/}{./figures/}}

\usepackage{xcolor}
\usepackage{bm}
\usepackage{braket}
\usepackage{hyperref}

\newcommand{\cmax}{C_{\mathrm{max}}}

\begin{document}

\title{Numerical Evidence for Exponential Speed-up of QAOA over Unstructured Search for Approximate Constrained Optimization}

\author{
    \IEEEauthorblockN{John Golden\IEEEauthorrefmark{1}\IEEEauthorrefmark{2}, 
    Andreas Bärtschi\IEEEauthorrefmark{1}\IEEEauthorrefmark{2}, 
    Daniel O'Malley\IEEEauthorrefmark{3},
    Stephan Eidenbenz\IEEEauthorrefmark{2}}
    \IEEEauthorblockA{\IEEEauthorrefmark{2}
    \textit{Information Sciences (CCS-3), Los Alamos National Laboratory},
	Los Alamos, NM 87544, USA
	}
    \IEEEauthorblockA{\IEEEauthorrefmark{3}
    \textit{Computational Earth Sciences (EES-16), Los Alamos National Laboratory},
	Los Alamos, NM 87544, USA
	}
    \IEEEauthorblockA{\IEEEauthorrefmark{1}
    Corresponding authors:
    \href{mailto:golden@lanl.gov}{golden@lanl.gov}, 
	\href{mailto:baertschi@lanl.gov}{baertschi@lanl.gov}
	}
}

\maketitle

\begin{abstract}
	Despite much recent work, the true promise and limitations of the Quantum Alternating Operator Ansatz (QAOA)~\cite{hadfield_qaoa} are unclear.
	A critical question regarding QAOA is to what extent its performance scales with the input size of the problem instance, 
	in particular the necessary growth in the number of QAOA rounds to reach a high approximation ratio.

	We present numerical evidence for an exponential speed-up of QAOA over Grover-style unstructured search~\cite{grover} 
	in finding approximate solutions to constrained optimization problems. Our result provides a strong hint that QAOA is
	able to exploit the structure of an optimization problem and thus overcome the lower bound for unstructured search~\cite{groverlower}. 
	
	To this end, we conduct a comprehensive numerical study on several Hamming-weight constrained optimization problems for which we include combinations of all 
	standardly studied mixer and phase separator Hamiltonians~\cite{cook2020kVC} (\emph{Ring} mixer, \emph{Clique} mixer, \emph{Obj}ective Value phase separator) 
	as well as quantum minimum-finding~\cite{DH96} inspired Hamiltonians~\cite{baertschi2020grover,golden2021thresholdbased} (\emph{Grover} mixer, \emph{Th}reshold-based phase separator).
	
	We identify \emph{Clique-Obj}-QAOA with an exponential speed-up over \emph{Grover-Th}-QAOA 
	and tie the latter's scaling to that of unstructured search, with all other QAOA combinations coming in at a distant third.
	Our result suggests that maximizing QAOA performance requires a judicious choice of mixer and phase separator,
	and should trigger further research into other QAOA variations.
\end{abstract}

\section{Introduction}\label{sec:intro}

\begin{figure*}
	\includegraphics[width=\linewidth]{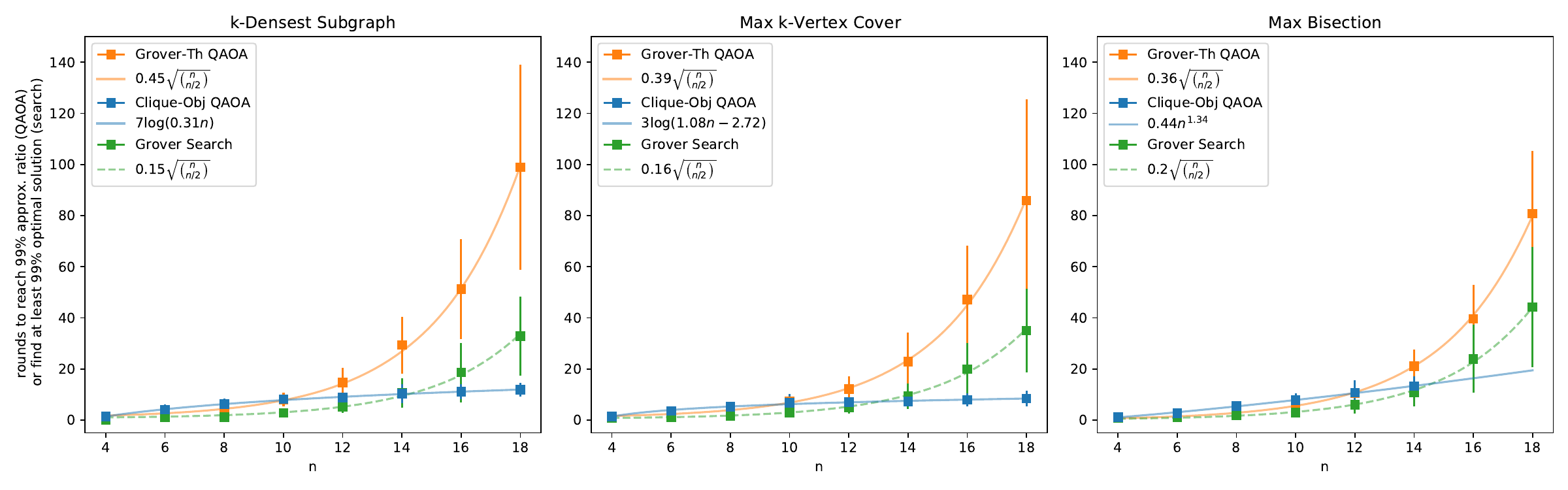}
	\caption{Peformance comparison for Clique-Obj (blue) and Grover-Th (orange) QAOA on $k$-Densest Subgraph and Maximum $k$-Vertex Cover from $n=4$ through $n=18$, and Max Bisection from $n=4$ through $n=14$. The $y$-axis shows the number of rounds of the respective QAOA which is sufficient to get a state with \textbf{approximation ratio 99\%}.
		For comparison we also show the number of rounds necessary for a Grover unstructured \emph{search} (green dashed) to sample with high probability a marked state of objective value at least $0.99 C_{\textrm{max}}-1$.\newline
		Square points represent average performance across 40 problem instances per $n$ on random Erd\"os-Renyi graphs $G(n,p=0.5)$, except for $n=4$ where we study all 64 possible four-vertex graphs. Vertical bars represent standard deviation across the problem instances.
		The smooth lines indicate the best functional fits to the data as given by weighted least squares. These results show that the Clique-Obj QAOA implementation significantly outperforms the Grover-Th QAOA, likely due to effective exploitation of underlying problem structure.	
	}
	\label{fig:all_n_fits}
\end{figure*}

The Quantum Alternating Operator Ansatz (QAOA) \cite{hadfield_qaoa} is a quantum-classical heuristic algorithm for finding approximately optimum solutions for combinatorial optimization problems.
Starting from an initial superposition of feasible solution states, QAOA repeatedly applies a phase separator operator followed by a mixer operator for a given number of rounds.
The phase separator gives a phase to solutions according to the objective function to be optimized, and the mixer operator generates interference amongst states in order to amplify high-quality solutions.
QAOA can be applied to a wide variety of combinatorial optimization problems by choosing an appropriate mixer and phase separator.

Analytical results regarding QAOA performance are rare due to the complexity of the mixer and phase separator Hamiltonians of even simple optimization problems, with the most well-known being a performance guarantee on MaxCut on 3-regular graphs~\cite{Farhi2014} over a small number of rounds~\cite{wurtz2021pg1,caha2022twisted}. 
For higher rounds or more complex problems, determining the optimal operators for each problem, and the amount of phasing and mixing for each round, generates a large number of parameters to tune.
Therefore much of the progress in QAOA research has come from numerical experimentation, treating it as more of a heuristic approach.
This is akin to classical techniques such as Simulated Annealing, Tabu Search, or Basin-hopping.
Such heuristics usually do not give performance guarantees, but can outperform provable methods in practice in terms of runtime and solution quality.

In the classical setting, the improved performance of heuristics over provable methods comes from better exploiting the structure of the problem being solved, but the nature of how this exploitation happens is difficult to capture in closed-form formulas.
Similarly, we ask whether QAOA can exploit problem structure. It is perhaps surprising that this question of whether QAOA successfully exploits problem structure has not been solved formally; there are strong opinions by leading figures on both sides (see e.g., https://scottaaronson.blog/?p=6457). 
A few results for unconstrained problems show that the depth of QAOA should grow logarithmically in the input size, in order for QAOA ``to see the whole graph'',~\cite{bravyi2020obstacles,farhi2020typical,farhi2020worstcase}, due to the non-entangling standard Transverse Field based mixer.

In a different approach~\cite{akshay2020reachability}, the standard QAOA (Transverse Field mixer, Objective value phase separator) on Max $k-$SAT was compared to the Grover Mixer and found to perform worse.
The phase separator, however, was in both cases the objective value phase separator, and despite interesting results on variational Grover search~\cite{akshay2020reachability,biamonte2018variational}, no direct connection was drawn between QAOA and unstructured search performance. 
We provide a clear example problem in this article showing that a QAOA variation indeed is able to exploit structure by exponentially outperforming unstructured search and thus overcoming the lower bound of unstructured search~\cite{groverlower}.

The recently introduced Grover mixer for constrained problems~\cite{baertschi2020grover} with threshold-based phase separator~\cite{golden2021thresholdbased} provides a natural
adaptation of unstructured search~\cite{grover} respectively its optimum-finding variant~\cite{DH96} into the QAOA language.
This \emph{Grover-Th} QAOA variation (discussed in more detail in Sections~\ref{sec:review}, \ref{sec:results}) effectively boosts amplitudes of all states with objective value above a given threshold. 
In contrast to a search, where one would like to maximize the probability to sample states above the threshold, Grover-Th-QAOA instead is optimized for the expectation value of sampled states, 
which depends on the two distributions of objective values above and below the threshold.
Thus the optimal choice of threshold to maximize QAOA performance depends on those distributions for the given problem instance as well as the number of rounds the QAOA will run.
However, the optimal angles for Grover-Th are generally all $\pi$ (except for the last round), regardless of problem, which is far simpler than other QAOA implementations.
Numerical evidence also shows that its scaling grows with the \emph{same asymptotic behaviour as an unstructured search}, with a fair comparison outlined in Section~\ref{sec:review}. 
Grover-Th is therefore a useful benchmark to compare other QAOA variations, and it is clear that a useful optimization scheme must, at a minimum, outperform this unstructured search approach.
We find this to be the case for a \emph{Clique-Obj} QAOA approach.

Our evidence is of course obtained through numerical simulation, which -- unlike its classical counterpart -- faces the problem of scalability as we can simulate only up to problem size 18, even using significant computational resources. We firmly believe that the asymptotic scaling of our QAOA variation becomes apparent (see blue line in Fig.~\ref{fig:all_n_fits}) by looking at instances from size 4 through 18 because there are no apparent constants or  lower-order terms in the runtime whose existence would suggest a comparatively better runtime at smaller instances than at truly large-scale instances. The tight confidence intervals and the high quality of the numerical fit further support our view that the fitted formula represents the true asymptotic runtime very closely. 
Nevertheless, we cannot formally exclude the possibility that asymptotic scaling is still hidden and we therefore use the terminology of ``evidence''' rather than ``numerical proof''.  

In this work we conduct a comprehensive numerical study of QAOA performance on several Hamming weight constrained optimization problems.
There are only three known mixers which work in this constrained context: the Ring, Clique,~\cite{angles-basin-nasa,cook2020kVC} and Grover mixers~\cite{baertschi2020grover}.
We pair each of these mixers with both the traditional objective value-based phase separator, as well as the newer threshold-based phased separator introduced in~\cite{golden2021thresholdbased}.
Our study tests QAOA performance on large number of random graphs of up to 18 vertices, which requires significant HPC resources despite our use of several problem-specific speed-up mechanisms; the high levels of entanglement inherent in these QAOA mixers render other speed-up tricks (such as \cite{lykov2020tensor}) largely ineffective.

This study across such a wide range of QAOA variations is the first of its kind in the QAOA literature, and has several important takeaways.
First, our main result is that the the Clique mixer with the objective value phase separator performs exponentially better than all other QAOA variations, including Grover-Th, see Fig.~\ref{fig:all_n_fits}.
Second, we link the exponential scaling of Grover-Th QAOA to that of Grover unstructured search.
Finally, none of the other QAOA variations consistently outperform Grover-Th. 
The wide variation of performance shows that the choice of mixer and phase separator play a significant, and previously underappreciated, role in QAOA performance.
Finally, this analysis is not intended to suggest that QAOA outperforms the best classical approaches -- we purposefully avoid classical comparisons.
Instead, we show that the ``easier'' task of outperforming unstructured search is, perhaps surprisingly, a non-trivial benchmark that numerous QAOA variations in the literature fail to pass.
Still, the existence of a QAOA variation that does surpass unstructured search strongly suggests that QAOA is able to exploit problem-specific structure. 
This should be viewed as a significant boost to the prospects of QAOA to someday achieve practical quantum advantage as a quantum optimization heuristic and trigger further research into QAOA variations.

\section{QAOA Review} \label{sec:review}
A QAOA algorithm begins with an objective function $C(x)$ encoding the optimization problem under consideration and a set of feasible solutions $S = \{x\}$.
One defines the cost Hamiltonian $H_C$ by $H_c\ket{x} = C(x)\ket{x}$.
The goal is to prepare a state $\ket{\psi}$ from which one can sample high-quality solutions. 

For the QAOA approaches considered here, we take our initial state as the uniform superposition of all feasible states, $\ket{\psi_0} = |S|^{-1/2}\sum_{x \in S} \ket{x}$.
For an unconstrained optimization problem of size $n$, $S$ is the set of all $n$-qubit computational basis states.
For constrained optimization, $S$ is restricted to some feasible subspace.

The quantum subroutine in QAOA consists of applying an alternating series of phase separator and mixer operators, defined in terms of Hamiltonians $H_P$ and $H_M$ respectively, on the initial state. 
The user inputs a number of rounds $p$ along with arrays of angles $\boldsymbol{\beta} = \{\beta_1,\ldots,\beta_p\}$, $\boldsymbol{\gamma} = \{\gamma_1,\ldots,\gamma_p\}$, from which a quantum computer prepares the state
\begin{equation}
	\ket{\boldsymbol{\beta}, \boldsymbol{\gamma}} =e^{-i\beta_p H_M}e^{-i \gamma_p H_P}\ldots e^{-i\beta_1 H_M}e^{-i \gamma_1 H_P}\ket{\psi_0}.
\end{equation}
Most commonly, a classical optimizer is used to tune the parameters $\boldsymbol{\beta}, \boldsymbol{\gamma}$ in order to maximize $\braket{H_C} := \braket{\boldsymbol{\beta}, \boldsymbol{\gamma}|H_C|\boldsymbol{\beta}, \boldsymbol{\gamma}}$~\cite{Farhi2014,hadfield_qaoa,qaoa-google},
but other cost functions have been suggested~\cite{barkoutsos2020cvar,li2020gibbs,larkin2020evaluation,bennett2021quantum}.
For problems with objective values in the range $0,\ldots,\cmax:=\max_{x\in S}{C(x)}$, there is an overhead polynomial in $\cmax$ to, with high probability, (i) get a sample above $\braket{H_C}-1$~\cite{Farhi2014} and (ii) estimate $\braket{H_C}$ within $\pm1$~\cite{cook2020kVC}.
Moving forward, we adopt the standard terminology of maximizing the approximation ratio, given by $\braket{H_C} / \cmax$~\cite{NPOcompendium,NPObook}. 

To begin the $i$-th round of the QAOA, the phase separator unitary applies a phase $e^{-i\gamma_i H_P}$ to each state, determined by the angle $\gamma_i$ as well as the phase separator Hamiltonian $H_P$.
The traditional definition of QAOA~\cite{hadfield_qaoa} uses an objective value-based phase separator, where each state is phased equal to the objective value for that state, $H_P = H_C$.

An alternative approach is to apply a phase only to states with an objective value above a threshold~\cite{golden2021thresholdbased}.
This threshold-based phase separator is defined for a given threshold parameter $th$ as
\begin{equation}\label{eq:phase-th}
  H_{Th}\ket{x} =
    \begin{cases}
	    0\ket{x} & \text{if } C(x) \leq  th,\\
      	1\ket{x} & \text{otherwise.}
    \end{cases}       
\end{equation}
This threshold is an additional parameter to be tuned to optimize QAOA performance.

After the phase separator unitary has applied a phase to each basis state, the mixing unitary $e^{-i\beta_i H_M}$ generates constructive and destructive interference between the states.
Here we consider three implementations of the mixing operator.
The Clique and Ring mixers~\cite{hadfield_qaoa}, $H_{Cl}$ and $H_R$, are given by
\begin{equation}
	H_{Cl} = \sum_{j>i} X_iX_j + Y_iY_j,\quad H_R = \sum_{j=i+1} X_iX_{j} + Y_i Y_{j}.\
\end{equation}
Here $X$ and $Y$ refer to the standard Pauli operators. 
The Clique mixer sums over all pairs of qubits, while the Ring mixer includes only adjacent qubits along a chain with periodic boundary. 
The Grover mixer $H_G$~\cite{baertschi2020grover,akshay2020reachability} is given by 
\begin{equation}
	H_G=\ket{\psi_0}\bra{\psi_0},
\end{equation}
where $\ket{\psi_0}$ denotes the equal superposition over all computational basis states of Hamming weight $k$, 
so-called Dicke states~\cite{baertschi2019dicke,baertschi2022shortdepth}.
These mixers are the only ones currently introduced in the literature which are applicable for Hamming weight $k$-constrained problems.
This is because they are the only mixers which only mix states of equal Hamming weight.
Mixers used for unconstrained problems, e.g. the Transverse Field mixer, are not applicable as they would mix with unfeasible solutions.

The QAOA framework works with any combination of these mixers and phase separator operators.
We refer to each choice by the name of the mixer followed by ``-Obj'' to refer to the objective value-based phase separator $H_C$, and ``-Th'' refers to the threshold-based phase separator $H_{Th}$.

Setting $\beta = \pi$ with the Grover mixer gives Grover's diffusion operator, $e^{-i \pi \ket{\psi_0}\bra{\psi_0}} = I - 2\ket{\psi_0}\bra{\psi_0}.$
Similarly, setting $\gamma = \pi$ with the threshold phase separator gives a phase of -1 to all states with objective value above the threshold and +1 to states below the threshold. 
Therefore the Grover-Th implementation QAOA with all angles set to $\pi$ is exactly equivalent to Grover search for states above a given threshold.\footnote{For a QAOA of $p$ rounds one must avoid the ``overshooting'' problem with Grover by setting $\beta=\gamma=0$ for all rounds after reaching a maximal expectation value.}
Furthermore, it was shown in~\cite{golden2021thresholdbased} that $\beta_i=\gamma_i=\pi$ are the optimal angles for Grover-Th.

\begin{figure*}[t]
	\includegraphics[width=\linewidth]{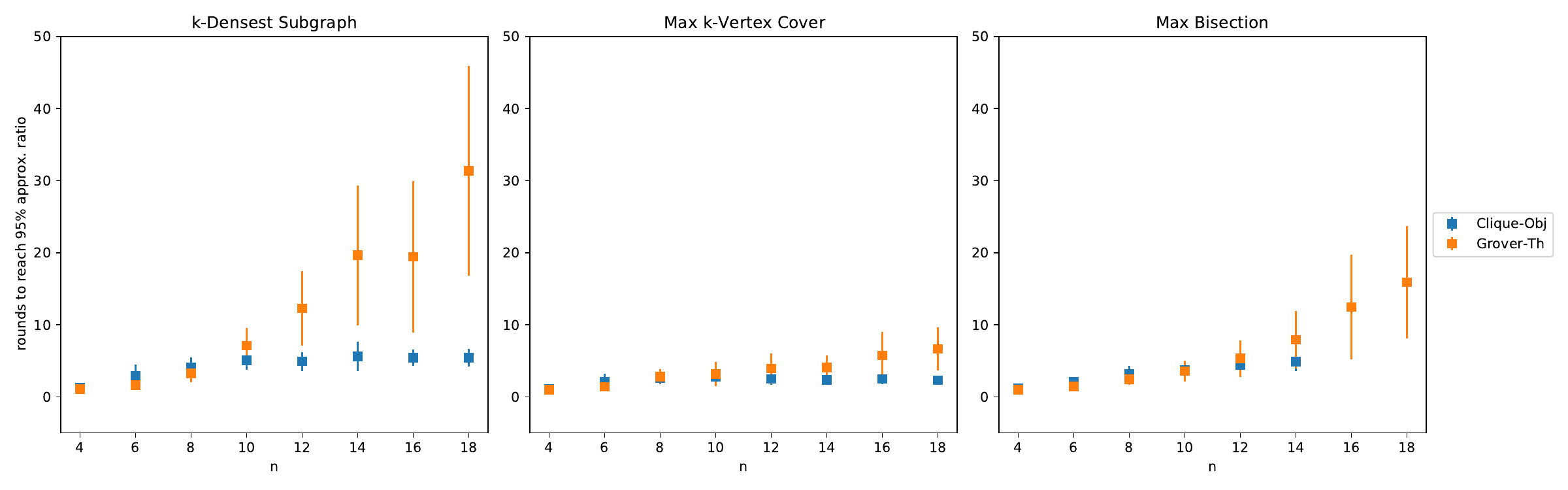}
	\caption{Peformance comparison for Clique-Obj (blue) and Grover-Th (orange) QAOA on $k$-Densest Subgraph and Maximum $k$-Vertex Cover from $n=4$ through $n=18$, and Max Bisection from $n=4$ through $n=14$. The $y$-axis shows the number of rounds of the respective QAOA which is sufficient to get a state of \textbf{approximation ratio 95\%}.\newline
		Square points represent average performance across 40 problem instances per $n$ on random Erd\"os-Renyi graphs $G(n,p=0.5)$, except for $n=4$ where we study all 64 possible four-vertex graphs. Vertical bars represent standard deviation across the problem instances.%
	}
	\label{fig:all_n_95}
\end{figure*}

While Grover-Th shares many similarities with the variational version of Grover's search algorithm, several critical differences remain.
Specifically, in applying Grover-Th to optimize the approximation ratio of the QAOA after a fixed number of rounds, the correct choice of threshold is non-trivial, see Fig.~\ref{fig:threshold_peaks} for an example.
Therefore Grover-Th, when used for approximate optimization, is not as simple as searching for some number of marked states.
This is also distinct from the Grover-like QAOA implementation of~\cite{zhang2017transverse} in that the circuit-level implementation is indistinguishable from Grover, and it is applicable to both constrained and unconstrained problems~\cite{akshay2020reachability,baertschi2020grover}.

However, since the asymptotic scaling of Grover's search algorithm is known, this provides a natural benchmark to first test Grover-Th QAOA and
then other QAOA variants against. 
In making this comparison, we use the number of rounds $p$ as a proxy for algorithmic runtime.
This ignores two additional sources of complexity: finding good angles and thresholds as well as the depth of the circuit-level implementations.
We discuss angle and threshold finding in Sec.~\ref{sec:discussion}.
All of these mixers and phase separators can be implemented exactly with polynomial-size circuits for the optimization problems we consider although their precise complexity is an active research topic~\cite{nasa2020XY,gu2021fast,baertschi2019dicke}. Here we restrict ourselves to a brief outline:
\begin{itemize}
	\item	\emph{Objective Value Phase Separator}: $H_C$ consists of Quadratic and (except for Max Bisection) Linear Pauli-Z terms, which pairwise commute, and hence can be implemented individually (e.g. with depth at most $O(n)$ using a SWAP network~\cite{swap-networks,qaoa-google}).
	\item	\emph{Threshold Phase Separator}: The cost $C(x)$ can be added into a Two's Complement~\cite{twos-complement} ancilla register, initialized with $-th-1$. Phasing can then be done simply by (inverse) phasing of the leading qubit which encodes the sign information, followed by an uncomputation of the register.
	\item	\emph{Ring Mixer}: The Jordan Wigner Transform maps the Ring Mixer Hamiltonian to a quadratic fermionic Hamiltonian~\cite{nasa2020XY}. 
		If $n$ is a power of two and the Hamming weight is odd, this Hamiltonian can be diagonalized using the 
		fermionic fast Fourier transformation~\cite{ferris2014fourier,verstraete2009quantum}, a method which can be extended if $n$ has only small prime factors~\cite{kivlichan2020improved}. 
		Otherwise, one can also use a Givens rotation network~\cite{kivlichan2018quantum}.
	\item	\emph{Clique Mixer}: The Clique Mixer, as a permutation-invariant qubit Hamiltonian~\cite{gu2021fast}, can be diagonalized using the Schur Transform~\cite{bacon2006efficient}. In particular, this results in phasing according to the total angular momentum.
	\item	\emph{Grover Mixer}: For Hamming-weight constrained problems, the Grover Mixer~\cite{baertschi2020grover} can be implemented with a $n$-controlled phase shift together with Dicke State preparation unitaries~\cite{baertschi2019dicke}. \\
\end{itemize}
An approximate implementation through Trotterization largely eliminates size differences, and so it is common in QAOA analysis to compare different mixers purely on a per-round basis~\cite{cook2020kVC,akshay2020reachability}. This is even more reasonable considering our results, which discriminate between exponential, polynomial and logarithmic scaling in the number of rounds necessary to reach high approximation ratios. 
However, a full analysis including the circuit depths leaves room for future work.

\section{Problems and Results}\label{sec:results}

We performed numerical simulations of all six combinations of mixers and phase separators -- Clique-Obj, Clique-Th, Ring-Obj, Ring-Th, Grover-Obj, and Grover-Th -- on three constrained optimization problems.
Of these combinations, only Clique-Obj consistently outperforms Grover-Th (over most rounds for non-small problem instances), see Figs.~\ref{fig:low-n},\ref{fig:all_n_fits}.

\subsection{Optimization Problems}
The problems we considered were $k$-Densest Subgraph, Maximum $k$-Vertex Cover, and Maximum Bisection.
These problems are defined on an undirected graph $G(V,E)$ with $|V|=n$.
$k$-Densest Subgraph is the problem of finding the subgraph with $k$ vertices that contains the most edges, while Max $k$-Vertex Cover is the problem of finding the $k$ vertices that cover the most edges.
Max Bisection is defined only for $k=n/2$ and asks for the partition of the graph into two equal subgraphs with the most edges crossing between partitions. 
If the adjacent vertices of $e\in E$ are labeled as $v^1_e$ and $v^2_e$, these problems are formally defined as 
\begin{equation}
	\text{maximize } \sum\nolimits_{e\in E} f(v^1_e \in V',v^2_e \in V')
\end{equation} 
over all $V'\subset V$ with $|V'|=k<n$, where $f=$ AND, OR, XOR 
for $k$-Densest Subgraph ($DS$), Max $k$-Vertex Cover ($VC$) and Max Bisection ($BS$) respectively.
Written in Pauli-Matrix terms, this translates to Cost Hamiltionians
\begin{align*}
	H_{C,DS}	& =	\frac{1}{4} \sum\nolimits_{\{u,v\}\in E} 1 \mathit{Id} + Z_uZ_v - Z_u - Z_v	\\	 
	H_{C,VC}	& =	\frac{1}{4} \sum\nolimits_{\{u,v\}\in E} 3 \mathit{Id} - Z_uZ_v - Z_u - Z_v	\\	 
	H_{C,BS}	& =	\frac{1}{2} \sum\nolimits_{\{u,v\}\in E} 1 \mathit{Id} - Z_uZ_v					 
\end{align*}
Note that the $H_{C,BS}$ is the same as the Maximum Cut Cost Hamiltonian~\cite{Farhi2014}; 
the restriction to the feasible subspace is done exclusively through the initial state $\ket{\psi_0}$ and the Mixer Hamiltonians, not through a penalty term in $H_C$.

The decision versions of our three problems are NP-hard for general graphs. The optimization variants considered here cover a broad spectrum in their known complexity results:
\begin{itemize}
	\item	Max Bisection is MaxSNP-complete~\cite{papadimitriou1991maxsnp} and the approximation hardness carries over from Max Cut, i.e. there is no polynomial-time approximation better than $\approx$0.941 unless $P=\mathit{NP}$~\cite{hastad2001inapproximability}, 
or better than $\approx$0.878 under the Unique Games Conjecture~\cite{khot2007ugc}.
The current best polynomial-time approximation ratio $0.8776$ is achieved with estimated runtime $O(n^{10^{100}})$~\cite{austrin2016bisection}.
	\item	For Max $k$-Vertex Cover, the corresponding known upper and lower bounds on polynomial-time approximation are $(1-\delta)$ for a small $\delta$~\cite[NP-hardness]{petrank1994hardness}, $0.944$~\cite[UGC-hardness]{austrin2011kvc,manurangsi2019}, $0.92$~\cite[approximation algorithm]{manurangsi2019}.
	\item	Similarly, $k$-Densest Subgraph has no PTAS (polynomial-time approximation scheme) under various complexity-theoretic assumptions~\cite{feige2002averagecase,khot2004noptas}, and the best known polynomial-time achievable approximation ratio is $n^{-1/4-\epsilon}$ for all $\epsilon>0$~\cite{kDSpoly}.
\end{itemize}

For these problems, the QAOA algorithm uses $n$ qubits and only computational basis states with Hamming weight $k$, i.e. exactly $k$ qubits set to $1$, represent feasible solutions.
This gives a total of $\binom{n}{k}$ feasible solutions, and the equal superposition of them is the Dicke state $\ket{D^n_k}$, which can be prepared in depths $\mathcal{O}(n)$ on linear nearest neighbor~\cite{baertschi2019dicke,aktar2022divideconquer}, 
and $\mathcal{O}(k \log \tfrac{n}{k})$ and $\mathcal{O}(k \surd \tfrac{n}{k})$ on grid and fully connected architectures, respectively~\cite{baertschi2022shortdepth}.
As discussed in Sec.~\ref{sec:review}, we focus on these constrained problems in connection to the Clique, Ring, Grover mixers which preserve Hamming weight and provide transitions between all feasible states.

Our performance metric in these comparisons is \emph{the number of rounds necessary to achieve an approximation ratio of 0.99} (we also give results for 0.95). 
With such a high approximation target, we can clearly observe in Fig.~\ref{fig:all_n_fits} the expected Grover-Th scaling of $\mathcal{O}(\binom{n}{k}^{1/2})$ as problem size increases.

For each problem class we evaluated the number of rounds necessary to reach .99 approximation ratio across 40 random instances for each $n$.
We then performed a weighted least squares fit of this data to both a logarithm ansatz, $a\log(bn + c)$ and a polynomial ansatz, $an^b + c$.
As seen in Fig.~\ref{fig:all_n_fits}, the logarithm ansatz provided a very close fit for $k$-Densest Subgraph and Max $k$-Vertex Cover, while Max Bisection could only be fit to the polynomial ansatz.
Both $k$-Densest Subgraph and Max $k$-Vertex Cover could also be fit to the polynomial ansatz, with exponent $\le 0.075$ at 95\% confidence level. 
Hence independent of the choice of fit, the data heavily suggests that for large graph problems, e.g. $10^{10}$ nodes, the mean number of rounds to achieve a high approximation ratio for these problems is $\mathcal{O}(100)$.
This belief is further supported by the lack of constants or lower-order terms in the runtime whose existence would suggest a comparatively better runtime at smaller instances than at truly large-scale instances.
The tight confidence intervals and the high quality of the numerical fit further support our view that the fitted formula represents the true asymptotic runtime very closely.

\subsection{Scaling Results}

As noted before in the beginning of Section~\ref{sec:review}, for problems with a maximum value $\cmax$, among $\cmax$ samples of $\ket{\boldsymbol{\beta}, \boldsymbol{\gamma}}$, there is at least one state of objective value $>\braket{H_C}-1$ with high probability $1-\frac{1}{m}$.
Hence, for comparison, we can also perform a Grover unstructured search for states which are marked if and only if their objective value lies above $\braket{H_C}-1$, and analyze the number of oracle calls which are necessary to measure one of these states with the high probability $1-\frac{1}{m}$.
We get the same asymptotic scaling (up to a constant pre-factor) as for Grover-Th QAOA, as illustrated in Figure~\ref{fig:all_n_fits}.

In Fig.~\ref{fig:all_n_95} we give results for Clique-Obj and Grover-Th with a target of 0.95 approximation ratio. 
This better captures the performance for truly approximate optimization, as opposed to the 0.99 approximation ratio, which for the problems under consideration is very close to optimal. 
In these results, the expected exponential scaling for Grover-Th is less obvious, due to the relatively large fraction and highly varied distribution of high-quality solutions in low-$n$-problem instances.
However, Clique-Obj starts to outperform Grover-Th at even smaller problem sizes than in the 99\% target approximation plots in Figure~\ref{fig:all_n_fits}.

\begin{figure*}
	\includegraphics[width=\linewidth]{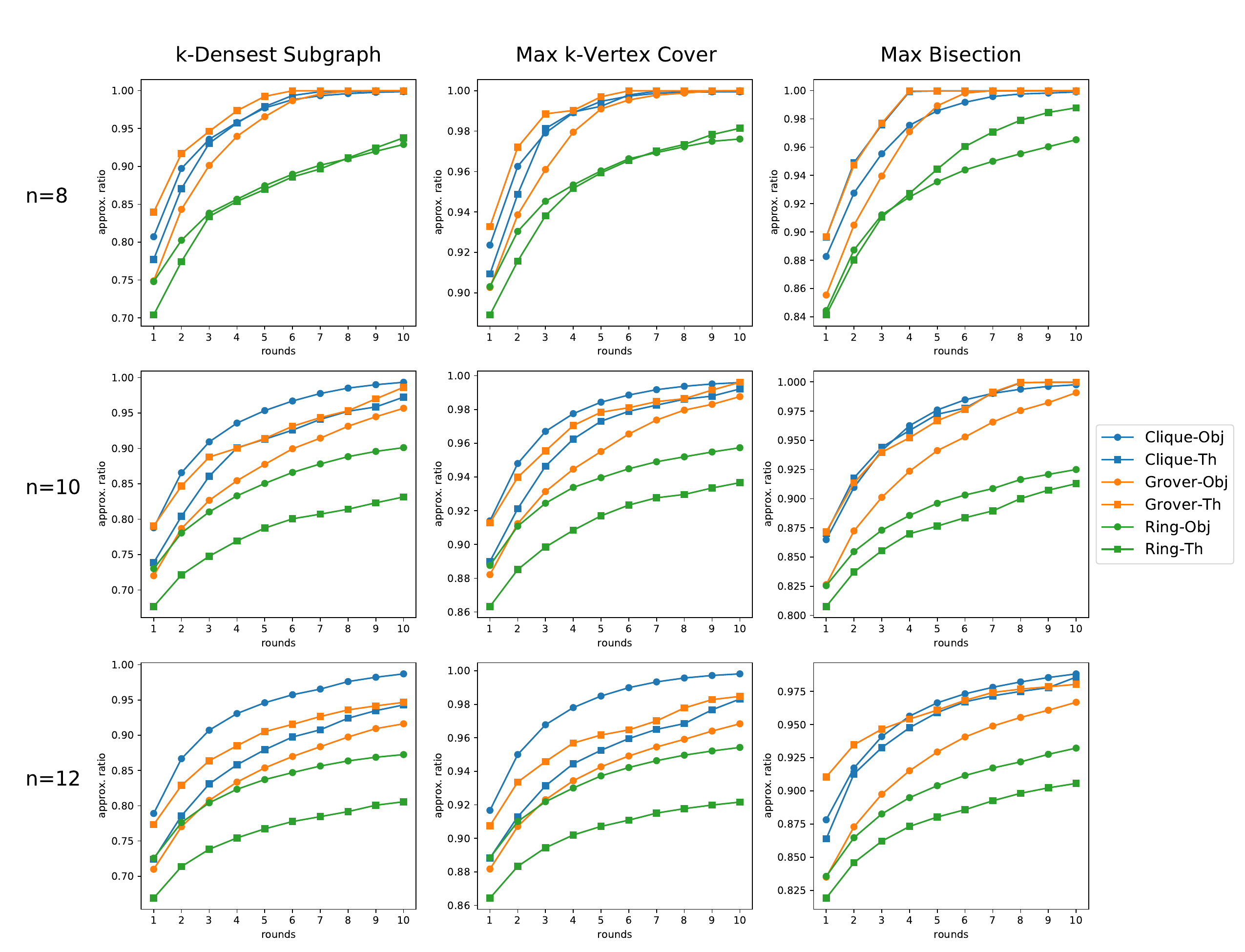}
	\caption{Round-by-round performance comparison for all QAOA implementations under consideration across $k$-Densest Subgraph, Max $k$-Vertex Cover, and Max Bisection for $n = 8,10,12$ and $k=n/2$, averaged over 12 problem instances on graphs of type $G(n,0.5)$.}
	\label{fig:low-n}
\end{figure*}

Looking at the rest of the mixer and phase separator combinations, we found that Grover-Th consistently outperformed the remaining QAOA operator choices -- Clique-Th, Grover-Obj, Ring-Th, and Ring-Obj.
We compared the mean performance over 12 random instances for each operator combination across the three problem classes at $n=8,10,12$ with $k=n/2$, see Fig.~\ref{fig:low-n}.
For $n=8$, Grover-Th is the best performing mixer for all problem classes, however by $n=10$ Clique-Obj has largely surpassed Grover-Th.
Meanwhile, the remaining mixers remain below Grover-Th, without a clear indication of an improvement as $n$ increases.
We give more details on the performance of these other approaches, and potential for future study, in Sec.~\ref{sec:discussion}.

\section{Discussion}\label{sec:discussion}

The approximation ratios of the different QAOA implementations generally followed the pattern Clique-Obj $\ge$ Grover-Th $\ge$ Clique-Th $\ge$ Grover-Obj $\ge$ Ring-Obj $\ge$ Ring-Th, though some difference in rankings is present at lower $n$.
For Clique and Ring mixers, the -Obj phase separator outperformed the -Th version.
This is in contrast to the Grover mixer, where Grover-Th always beat Grover-Obj (consistent with findings from~\cite{golden2021thresholdbased}, which extend up to $n=40$).
Clique-Th performed almost identically to Grover-Th for Max Bisection, but was worse for $k$-Densest Subgraph and Max $k$-Vertex Cover.

The observation that Clique-Obj significantly outperforms Grover-Th is dependent on two important elements: robust angle \& threshold finding and large problem size.

\begin{figure*}
	\includegraphics[width=\linewidth]{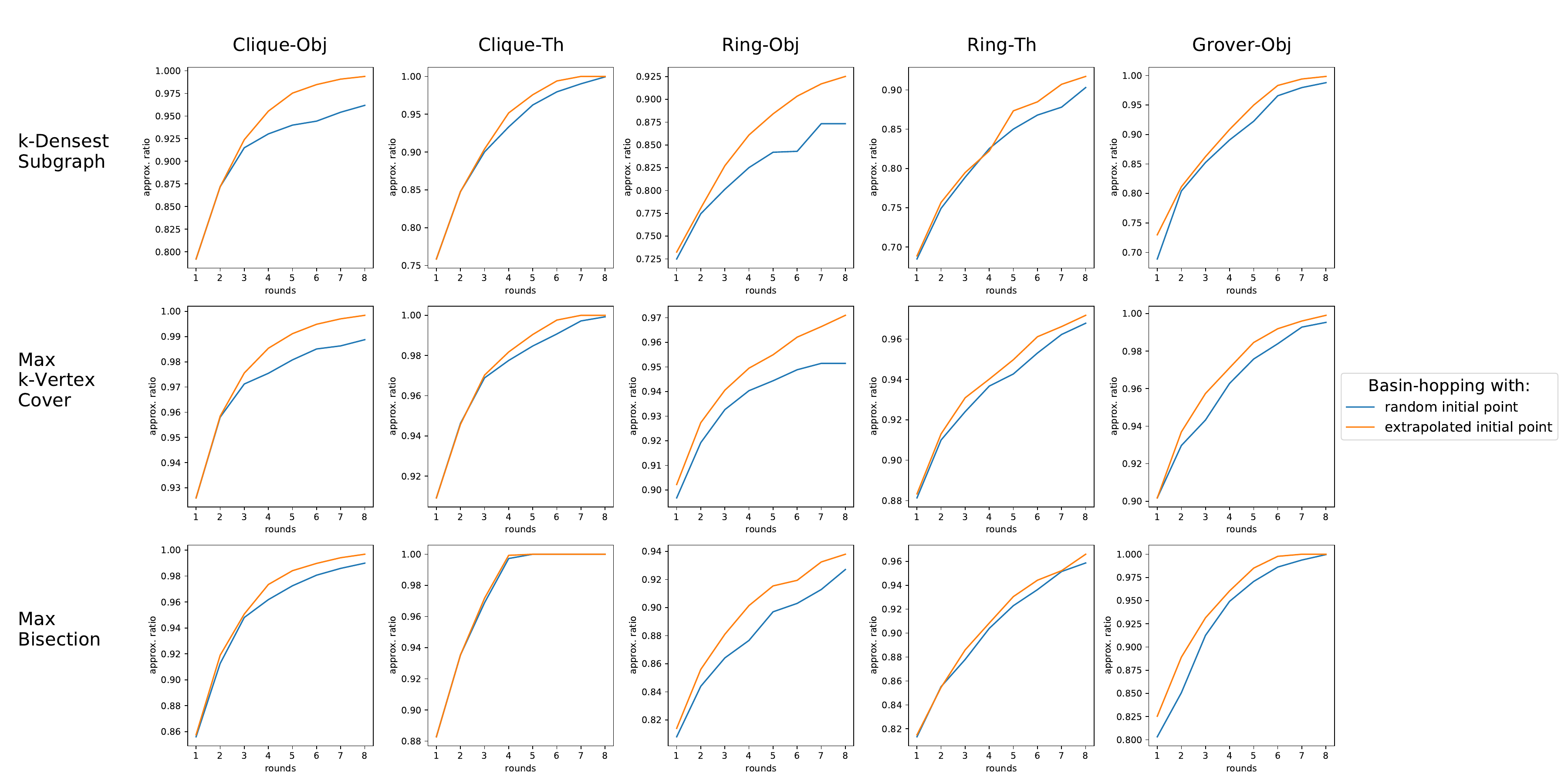}
	\caption{Comparing angle finding with basin-hopping starting at random initial point versus an extrapolated initial point. 
	This is averaging results over 12 random $n=8, k=4$ $k$-Densest Subgraph problems, and 100 basin-hopping iterations per round.}
	\label{fig:angle_finding}
\end{figure*}

\subsection{Angle \& Threshold Finding} 
\begin{figure}
\includegraphics[width=\linewidth]{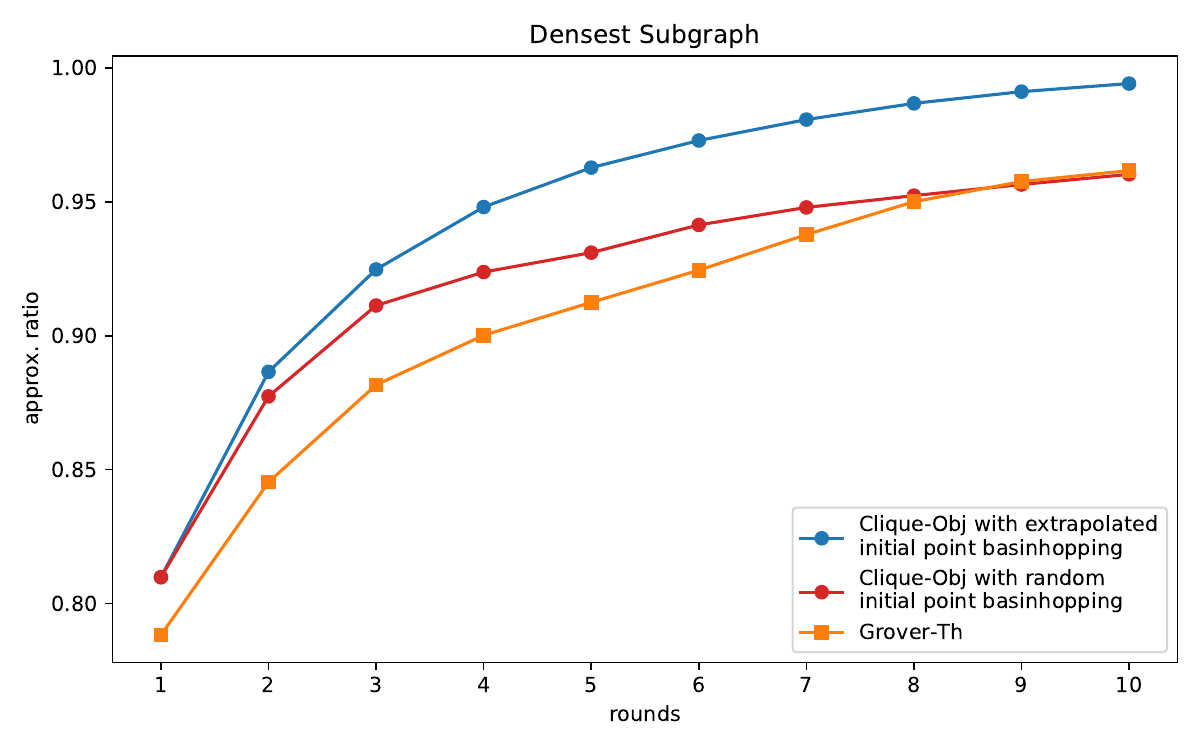}
\caption{Relative performance difference for Clique-Obj with basin-hopping from an extrapolated initial point versus a random initial point, averaged over 12 $k$-Densest Subgraph problem instances with $n=12, k=6$. The critical fact that Clique-Obj outperforms Grover-Th at high rounds is only apparent when using high quality angle-finding.}
\label{fig:af-n12comp}
\end{figure}

Our initial studies employed the traditional basin-hopping approach~\cite{nasa2020XY}, which uses a random initial collection of angles to begin optimizing for a $p$-round QAOA.
However we observed improved performance by using an inductive approach, partially inspired by~\cite{cook2020kVC}.
Specifically, for a given problem we take the optimal angles for the $(p-1)$-round QAOA, $\bm{\beta}_{p-1} = \{\beta_1,\ldots,\beta_{p-1}\}, \bm{\gamma}_{p-1} = \{\gamma_1,\ldots,\gamma_{p-1}\}$, and begin our basin-hopping search for angles for the $p$-round QAOA with
\begin{equation}\label{eq:extrap-angles}
	\bm{\beta}_p = \{\beta_1,\ldots,\beta_{p-1},\beta_{p-1}\},~ \bm{\gamma}_p = \{\gamma_1,\ldots,\gamma_{p-1}, \gamma_{p-1}\}.
\end{equation}
See Fig.~\ref{fig:angle_finding} for results comparing the extrapolated basin-hopping angle-finding scheme against basin-hopping from a random initial point at $n=8$.
This improved angle finding becomes particularly important for $n\ge12$,
as Clique-Obj with our extrapolated basin-hopping approach outperforms Grover-Th, whereas Clique-Obj with random basin-hopping does not, see Fig.~\ref{fig:af-n12comp}.
This result emphasizes that high-quality angle-finding can significantly change the performance of QAOA, particularly for large number of rounds.

For the specific case of Clique-Obj on $k$-Densest Subgraph and Max $k$-Vertex Cover, we observed over thousands of examples that the basin-hopping algorithm never left the local optimum it arrived at after starting at the extrapolated point.
Therefore, in order to accelerate our analysis we found the angles for all $p$ rounds by using Gradient Descent from Eq.~(\ref{eq:extrap-angles}).

Due to the optimality of Grover's algorithm for unstructured search, and the similar scaling of Grover-Th QAOA,
the performance gains from Clique-Obj must come from exploiting the structure of each individual problem instance via angle-finding.
A common concern for asymptotic QAOA performance is the complexity of finding good angles.
These results show that relatively simple and low-complexity angle-finding heuristics can exist for problems of interest up to roughly 30 rounds.
There are several additional sources of overhead which in our experience did not significantly contribute: finding angles for rounds $1,\ldots,p-1$ before the angles for round $p$ can be found, the cost of Gradient Descent for each round, and the cost of finding the optimal threshold for Grover-Th for a given round. 

\begin{figure*}
	\includegraphics[width=\linewidth]{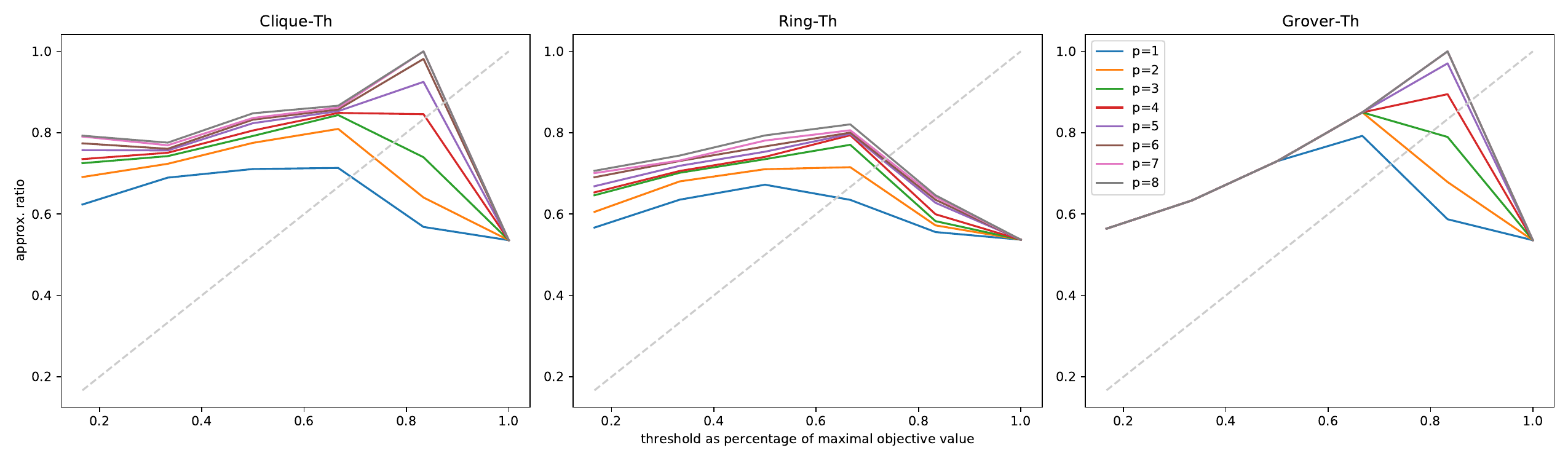}
	\caption{Approximation ratio as a function of threshold for Clique-Th, Ring-Th, and Grover-Th over 8 rounds on a single $n=8, k=4$ $k$-Densest Subgraph problem. The optimal choice of threshold to maximize approximation ratio is a highly non-trivial function of the number of rounds of the QAOA, and the resulting approximation ratio is greater than the threshold.}
	\label{fig:threshold_peaks}
\end{figure*}

Threshold-based phase separators require an algorithm for determining the optimal threshold for a given number of rounds $p$.
The simplest approach is to try every possible threshold value.
This is generally not too computationally onerous as constrained optimization problems, such as those considered in this work, have a number of distinct objective values which grows at-worst polynomially with problem size. 
For example, $k$-Densest Subgraph, Max $k$-Vertex Cover, and Max Bisection can have at most $\mathcal{O}(n^2)$ distinct objective values.
However, robust numerical experimentation points to the several improvements that can be made over this brute force search.
As an example see Fig.~\ref{fig:threshold_peaks}, where we plot approximation ratio as a function of threshold for a single $n=8,k=4$ $k$-Densest Subgraph problem over 8 rounds.
This example showcases several salient points which we can use to improve our threshold-finding. 

First, for all mixers the threshold which produces the highest approximation ratio for a given round $p$ is always greater than or equal to the threshold which produced the highest approximation ratio for round $p-1$. 
In practice, this heavily constrains the search space as even for $p=1$ the threshold which produces the highest approximation ratio tends to be reasonably large.

Second, with Grover-Th, for a given round the approximation ratio as a function of threshold monotonically increases up to a peak value, then monotonically decreases.
This peak structure allows for a modified binary search, as recently introduced~\cite{golden2021thresholdbased}.

Third, Clique-Th and Ring-Th do not necessarily follow this peak structure, therefore at round $p$ one must conduct an exhaustive search of all threshold values above the optimal $p-1$ threshold value.

\subsection{Problem Size}
The simplified angle-finding heuristic described above reduced our simulation time by a factor of 100 for Clique-Obj on $k$-Densest Subgraph and Max $k$-Vertex Cover, which allowed us to analyze problems up to $n=18$ and $p\approx30$, with the largest problem instances taking $\mathcal{O}(\text{1 day})$ compute time on an NVIDIA RTX A6000 with 48GB of memory.
For Max Bisection we required the full basin-hopping approach to get good results, and thus were only able to extend our results to $n=14$.
Due to the highly entangling nature of the mixers under consideration, we had to employ full statevector simulation of the QAOA algorithms and could not employ simplifications~\cite{lykov2020tensor} or the Schur transform (which needs polylogarithmically many ancilla qubits) to reduce simulation overhead. 
Analyzing problems of this size proved critical, as the improved performance of Clique-Obj is only evident for $n\ge12$.
This emphasizes the need for future study at large $n$.

In this work we have only studied random graphs with edge probability = 0.5 and we have restricted to the cases $k=n/2$ for $k$-Densest Subgraph and Max $k$-Vertex Cover.
Additional analysis is necessary to see if the relative QAOA implementation performance persists in other graph types, particularly since dense graphs often admit a classical PTAS~\cite{arora1999denseptas}.
However, we note that in~\cite{golden2021thresholdbased} Grover-Obj was compared against Grover-Th on the same problem classes as well as Max Cut, with differing edge probabilites and $k$ values. 
In these comparisons there was little variation in relative performance (with Grover-Th always outperforming Grover-Obj) across this more diverse set of problems.

The results for Clique-Th, Ring-Th, and Ring-Obj for $n\le12$ do not suggest a high likelihood of outperforming Grover-Th, however we cannot rule this possibility out.
For example, we see in Fig.~\ref{fig:low-n} that in the final round of Max Bisection results at $n=12$, Clique-Th outperformed Grover-Th.
Extending these results to even higher $n$, or discovering improved angle-finding techniques, may show further relative performance gains.

\subsection{Other QAOA Variants}

In this paper, we have shown the importance of numerically investigating the scaling of QAOA with growing input size:

On one hand, we have observed that the ranking among different QAOA variants may change once we go beyond small problem sizes, e.g., $\binom{10}{5}\approx 250$ feasible states, as shown in Figure~\ref{fig:low-n}.
In a more recent, related paper~\cite{golden2023quantum}, we have made similar observations for unconstrained problems: 
An extensive study of different mixers as well as the objective value and threshold phase separators had Grover-Th outperform other combinations
up to problems with $2^8 \approx 250$ feasible states, before dropping below a select few other QAOA variants. 

On the other hand, we do not expect QAOA to reach very high approximation ratios for growing input sizes when keeping the number of rounds constant. 
Given the large amount of computational resources spent on different initial angle finding strategies for $k$-Densest Subgraph, and Maximum $k$-Vertex Cover, 
as well as the small deviations across different input graphs, we are confident in the logarithmic scaling in the number of rounds being necessary 
to maintain the approximation ratio of 99\%. 
We believe it will be important to further investigate the \emph{asymptotic scaling} of QAOA, studying the \emph{tradeoffs} between \emph{input size} $n$, 
number of \emph{rounds} $p$, and \emph{approximation ratio} and/or \emph{success probabilities}: 
Such a related approach was recently also studied~\cite{boulebnane2022solving} on QAOA for a constraint satisfaction problem (as opposed to optimization):
They numerically investigated the asymptotic average success probability for random $8$-SAT problems with growing input size $n$,
for fixed numbers of rounds up to $p=60$.

\section{Acknowledgements}

We thank Satyajayant Misra for comments on this manuscript. This material is based upon work supported by the U.S. Department of Energy, Office of Science, National Quantum Information Science Research Centers, Quantum Science Center. We also thank the Laboratory Directed Research and Development Program at Los Alamos National Laboratory for support.\hfill LA-UR 22-20645.

\bibliographystyle{plainurl}
\bibliography{qaoa_comp}

\begin{thebibliography}{10}

\bibitem{akshay2020reachability}
V.~Akshay, H.~Philathong, M.~E.~S. Morales, and J.~D. Biamonte.
\newblock {Reachability Deficits in Quantum Approximate Optimization}.
\newblock {\em Physical Review Letters}, 124(9):090504, 2020.
\newblock \href {http://arxiv.org/abs/1906.11259} {\path{arXiv:1906.11259}},
  \href {https://doi.org/10.1103/PhysRevLett.124.090504}
  {\path{doi:10.1103/PhysRevLett.124.090504}}.

\bibitem{aktar2022divideconquer}
Shamminuj Aktar, Andreas B{\"{a}}rtschi, Abdel-Hameed~A. Badawy, and Stephan
  Eidenbenz.
\newblock {A Divide-and-Conquer Approach to Dicke State Preparation}.
\newblock {\em IEEE Transactions on Quantum Engineering}, 3:3101816, May 2022.
\newblock \href {http://arxiv.org/abs/2112.12435} {\path{arXiv:2112.12435}},
  \href {https://doi.org/10.1109/TQE.2022.3174547}
  {\path{doi:10.1109/TQE.2022.3174547}}.

\bibitem{arora1999denseptas}
Sanjeev Arora, David Karger, and Marek Karpinski.
\newblock {Polynomial Time Approximation Schemes for Dense Instances of NP-Hard
  Problems}.
\newblock {\em Journal of Computer and System Sciences}, 58(1):193--210, 1999.
\newblock \href {https://doi.org/10.1006/jcss.1998.1605}
  {\path{doi:10.1006/jcss.1998.1605}}.

\bibitem{NPObook}
G.~Ausiello, P.~Crescenzi, G.~Gambosi, V.~Kann, A.~Marchetti-Spaccamela, and
  M.~Protasi.
\newblock {\em {Complexity and Approximation}}.
\newblock Springer, November 1999.
\newblock \href {https://doi.org/10.1007/978-3-642-58412-1}
  {\path{doi:10.1007/978-3-642-58412-1}}.

\bibitem{austrin2016bisection}
Per Austrin, Siavosh Benabbas, and Konstantinos Georgiou.
\newblock {Better Balance by Being Biased: A 0.8776-Approximation for Max
  Bisection}.
\newblock {\em ACM Transactions on Algorithms}, 13(1), 2016.
\newblock \href {http://arxiv.org/abs/1205.0458} {\path{arXiv:1205.0458}},
  \href {https://doi.org/10.1145/2907052} {\path{doi:10.1145/2907052}}.

\bibitem{austrin2011kvc}
Per Austrin, Subhash Khot, and Muli Safra.
\newblock {Inapproximability of Vertex Cover and Independent Set in Bounded
  Degree Graphs}.
\newblock {\em Theory of Computing}, 7(3):27--43, 2011.
\newblock \href {https://doi.org/10.4086/toc.2011.v007a003}
  {\path{doi:10.4086/toc.2011.v007a003}}.

\bibitem{bacon2006efficient}
Dave Bacon, Isaac~L Chuang, and Aram~W Harrow.
\newblock {Efficient quantum circuits for Schur and Clebsch-Gordan transforms}.
\newblock {\em Physical Review Letters}, 97(17):170502, 2006.
\newblock \href {http://arxiv.org/abs/quant-ph/0407082}
  {\path{arXiv:quant-ph/0407082}}, \href
  {https://doi.org/10.1103/PhysRevLett.97.170502}
  {\path{doi:10.1103/PhysRevLett.97.170502}}.

\bibitem{barkoutsos2020cvar}
Panagiotis~Kl. Barkoutsos, Giacomo Nannicini, Anton Robert, Ivano Tavernelli,
  and Stefan Woerner.
\newblock {Improving Variational Quantum Optimization using CVaR}.
\newblock {\em Quantum}, 4:256, 2020.
\newblock \href {http://arxiv.org/abs/1907.04769} {\path{arXiv:1907.04769}},
  \href {https://doi.org/10.22331/q-2020-04-20-256}
  {\path{doi:10.22331/q-2020-04-20-256}}.

\bibitem{baertschi2019dicke}
Andreas B{\"a}rtschi and Stephan Eidenbenz.
\newblock {Deterministic Preparation of Dicke States}.
\newblock In {\em 22nd International Symposium on Fundamentals of Computation
  Theory FCT'19}, pages 126--139, 2019.
\newblock \href {http://arxiv.org/abs/1904.07358} {\path{arXiv:1904.07358}},
  \href {https://doi.org/10.1007/978-3-030-25027-0_9}
  {\path{doi:10.1007/978-3-030-25027-0_9}}.

\bibitem{baertschi2022shortdepth}
Andreas B{\"{a}}rtschi and Stephan Eidenbenz.
\newblock {Short-Depth Circuits for Dicke State Preparation}.
\newblock In {\em IEEE International Conference on Quantum Computing \&
  Engineering QCE'22}, pages 87--96, September 2022.
\newblock \href {http://arxiv.org/abs/2207.09998} {\path{arXiv:2207.09998}},
  \href {https://doi.org/10.1109/QCE53715.2022.00027}
  {\path{doi:10.1109/QCE53715.2022.00027}}.

\bibitem{groverlower}
Charles~H. Bennett, Ethan Bernstein, Gilles Brassard, and Umesh Vazirani.
\newblock {Strengths and Weaknesses of Quantum Computing}.
\newblock {\em SIAM Journal on Computing}, 26(5):1510--1523, 1997.
\newblock \href {http://arxiv.org/abs/quant-ph/9701001}
  {\path{arXiv:quant-ph/9701001}}, \href
  {https://doi.org/10.1137/S0097539796300933}
  {\path{doi:10.1137/S0097539796300933}}.

\bibitem{bennett2021quantum}
Tavis Bennett and Jingbo~B Wang.
\newblock Quantum optimisation via maximally amplified states.
\newblock {\em arXiv e-prints}, 2021.
\newblock \href {http://arxiv.org/abs/2111.00796} {\path{arXiv:2111.00796}}.

\bibitem{kDSpoly}
Aditya Bhaskara, Moses Charikar, Eden Chlamtac, Uriel Feige, and Aravindan
  Vijayaraghavan.
\newblock {Detecting High Log-Densities: An $\mathcal{O}(n^{1/4})$
  Approximation for Densest k-Subgraph}.
\newblock In {\em 42nd ACM Symposium on Theory of Computing STOC'10}, pages
  201--210, 2010.
\newblock \href {http://arxiv.org/abs/1001.2891} {\path{arXiv:1001.2891}},
  \href {https://doi.org/10.1145/1806689.1806719}
  {\path{doi:10.1145/1806689.1806719}}.

\bibitem{boulebnane2022solving}
Sami Boulebnane and Ashley Montanaro.
\newblock Solving boolean satisfiability problems with the quantum approximate
  optimization algorithm.
\newblock {\em arXiv preprint}, 2022.
\newblock \href {http://arxiv.org/abs/2208.06909} {\path{arXiv:2208.06909}}.

\bibitem{bravyi2020obstacles}
Sergey Bravyi, Alexander Kliesch, Robert Koenig, and Eugene Tang.
\newblock Obstacles to variational quantum optimization from symmetry
  protection.
\newblock {\em Physical Review Letters}, 125:260505, 2020.
\newblock \href {http://arxiv.org/abs/1910.08980} {\path{arXiv:1910.08980}},
  \href {https://doi.org/10.1103/PhysRevLett.125.260505}
  {\path{doi:10.1103/PhysRevLett.125.260505}}.

\bibitem{baertschi2020grover}
Andreas Bärtschi and Stephan Eidenbenz.
\newblock {Grover Mixers for QAOA: Shifting Complexity from Mixer Design to
  State Preparation}.
\newblock In {\em IEEE International Conference on Quantum Computing \&
  Engineering QCE'20}, pages 72--82, 2020.
\newblock \href {http://arxiv.org/abs/2006.00354} {\path{arXiv:2006.00354}},
  \href {https://doi.org/10.1109/QCE49297.2020.00020}
  {\path{doi:10.1109/QCE49297.2020.00020}}.

\bibitem{caha2022twisted}
Libor Caha, Alexander Kliesch, and Robert Koenig.
\newblock Twisted hybrid algorithms for combinatorial optimization.
\newblock {\em arXiv e-prints}, 2022.
\newblock \href {http://arxiv.org/abs/2203.00717} {\path{arXiv:2203.00717}}.

\bibitem{cook2020kVC}
Jeremy Cook, Stephan Eidenbenz, and Andreas B{\"a}rtschi.
\newblock {The Quantum Alternating Operator Ansatz on Maximum k-Vertex Cover}.
\newblock In {\em IEEE International Conference on Quantum Computing \&
  Engineering QCE'20}, pages 83--92, 2020.
\newblock \href {http://arxiv.org/abs/1910.13483} {\path{arXiv:1910.13483}},
  \href {https://doi.org/10.1109/QCE49297.2020.00021}
  {\path{doi:10.1109/QCE49297.2020.00021}}.

\bibitem{NPOcompendium}
Pierluigi Crescenzi and Viggo Kann.
\newblock A compendium of {NP} optimization problems.
\newblock URL: \url{https://www.csc.kth.se/tcs/compendium/}.

\bibitem{DH96}
Christoph Durr and Peter Hoyer.
\newblock A quantum algorithm for finding the minimum.
\newblock {\em arXiv e-prints}, 1996.
\newblock \href {http://arxiv.org/abs/quant-ph/9607014}
  {\path{arXiv:quant-ph/9607014}}.

\bibitem{farhi2020typical}
Edward Farhi, David Gamarnik, and Sam Gutmann.
\newblock {The quantum approximate optimization algorithm needs to see the
  whole graph: A typical case}.
\newblock {\em arXiv e-prints}, 2020.
\newblock \href {http://arxiv.org/abs/2004.09002} {\path{arXiv:2004.09002}}.

\bibitem{farhi2020worstcase}
Edward Farhi, David Gamarnik, and Sam Gutmann.
\newblock {The quantum approximate optimization algorithm needs to see the
  whole graph: Worst case examples}.
\newblock {\em arXiv e-prints}, 2020.
\newblock \href {http://arxiv.org/abs/2005.08747} {\path{arXiv:2005.08747}}.

\bibitem{Farhi2014}
Edward Farhi, Jeffrey Goldstone, and Sam Gutmann.
\newblock {A Quantum Approximate Optimization Algorithm}.
\newblock {\em arXiv e-prints}, 2014.
\newblock \href {http://arxiv.org/abs/1411.4028} {\path{arXiv:1411.4028}}.

\bibitem{feige2002averagecase}
Uriel Feige.
\newblock {Relations between Average Case Complexity and Approximation
  Complexity}.
\newblock In {\em 34th ACM Symposium on Theory of Computing STOC'02}, pages
  534--543, 2002.
\newblock \href {https://doi.org/10.1145/509907.509985}
  {\path{doi:10.1145/509907.509985}}.

\bibitem{ferris2014fourier}
Andrew~J Ferris.
\newblock Fourier transform for fermionic systems and the spectral tensor
  network.
\newblock {\em Physical Review Letters}, 113(1):010401, 2014.
\newblock \href {http://arxiv.org/abs/1310.7605} {\path{arXiv:1310.7605}},
  \href {https://doi.org/10.1103/PhysRevLett.113.010401}
  {\path{doi:10.1103/PhysRevLett.113.010401}}.

\bibitem{golden2023quantum}
John Golden, Andreas B{\"a}rtschi, Daniel O'Malley, and Stephan Eidenbenz.
\newblock {The Quantum Alternating Operator Ansatz for Satisfiability
  Problems}.
\newblock {\em arXiv preprint}, 2023.
\newblock \href {http://arxiv.org/abs/2301.11292} {\path{arXiv:2301.11292}}.

\bibitem{golden2021thresholdbased}
John Golden, Andreas Bärtschi, Daniel O'Malley, and Stephan Eidenbenz.
\newblock {Threshold-Based Quantum Optimization}.
\newblock In {\em IEEE International Conference on Quantum Computing \&
  Engineering QCE'21}, pages 137--147, 2021.
\newblock \href {http://arxiv.org/abs/2106.13860} {\path{arXiv:2106.13860}},
  \href {https://doi.org/10.1109/QCE52317.2021.00030}
  {\path{doi:10.1109/QCE52317.2021.00030}}.

\bibitem{grover}
Lov~K. Grover.
\newblock {A Fast Quantum Mechanical Algorithm for Database Search}.
\newblock In {\em 28th Annual ACM Symposium on Theory of Computing STOC'96},
  pages 212--219, 1996.
\newblock \href {http://arxiv.org/abs/quant-ph/9605043}
  {\path{arXiv:quant-ph/9605043}}, \href
  {https://doi.org/10.1145/237814.237866} {\path{doi:10.1145/237814.237866}}.

\bibitem{gu2021fast}
Shouzhen Gu, Rolando~D Somma, and Burak {\c{S}}ahino{\u{g}}lu.
\newblock Fast-forwarding quantum evolution.
\newblock {\em Quantum}, 5:577, 2021.
\newblock \href {http://arxiv.org/abs/2105.07304} {\path{arXiv:2105.07304}},
  \href {https://doi.org/10.22331/q-2021-11-15-577}
  {\path{doi:10.22331/q-2021-11-15-577}}.

\bibitem{hadfield_qaoa}
Stuart Hadfield, Zhihui Wang, Bryan O’Gorman, Eleanor~G Rieffel, Davide
  Venturelli, and Rupak Biswas.
\newblock From the quantum approximate optimization algorithm to a quantum
  alternating operator ansatz.
\newblock {\em Algorithms}, 12(2):34, 2019.
\newblock \href {http://arxiv.org/abs/1709.03489} {\path{arXiv:1709.03489}},
  \href {https://doi.org/10.3390/a12020034} {\path{doi:10.3390/a12020034}}.

\bibitem{qaoa-google}
Matthew~P. Harrigan et~al.
\newblock Quantum approximate optimization of non-planar graph problems on a
  planar superconducting processor.
\newblock {\em Nature Physics}, 17(3):332--336, 2021.
\newblock \href {http://arxiv.org/abs/2004.04197} {\path{arXiv:2004.04197}},
  \href {https://doi.org/10.1038/s41567-020-01105-y}
  {\path{doi:10.1038/s41567-020-01105-y}}.

\bibitem{hastad2001inapproximability}
Johan H\r{a}stad.
\newblock {Some Optimal Inapproximability Results}.
\newblock {\em Journal of the ACM}, 48(4):798–--859, 2001.
\newblock \href {https://doi.org/10.1145/502090.502098}
  {\path{doi:10.1145/502090.502098}}.

\bibitem{zhang2017transverse}
Zhang Jiang, Eleanor~G. Rieffel, and Zhihui Wang.
\newblock Near-optimal quantum circuit for grover's unstructured search using a
  transverse field.
\newblock {\em Physical Review A}, 95:062317, Jun 2017.
\newblock \href {http://arxiv.org/abs/1702.02577} {\path{arXiv:1702.02577}},
  \href {https://doi.org/10.1103/PhysRevA.95.062317}
  {\path{doi:10.1103/PhysRevA.95.062317}}.

\bibitem{khot2004noptas}
S.~Khot.
\newblock {Ruling out PTAS for graph min-bisection, densest subgraph and
  bipartite clique}.
\newblock In {\em 45th IEEE Symposium on Foundations of Computer Science
  FOCS'04}, pages 136--145, 2004.
\newblock \href {https://doi.org/10.1109/FOCS.2004.59}
  {\path{doi:10.1109/FOCS.2004.59}}.

\bibitem{khot2007ugc}
Subhash Khot, Guy Kindler, Elchanan Mossel, and Ryan O'Donnell.
\newblock {Optimal Inapproximability Results for MAX‐CUT and Other
  2‐Variable CSPs?}
\newblock {\em SIAM Journal on Computing}, 37(1):319--357, 2007.
\newblock \href {https://doi.org/10.1137/S0097539705447372}
  {\path{doi:10.1137/S0097539705447372}}.

\bibitem{kivlichan2020improved}
Ian~D Kivlichan, Craig Gidney, Dominic~W Berry, Nathan Wiebe, Jarrod McClean,
  Wei Sun, Zhang Jiang, Nicholas Rubin, Austin Fowler, Al{\'a}n Aspuru-Guzik,
  et~al.
\newblock Improved fault-tolerant quantum simulation of condensed-phase
  correlated electrons via trotterization.
\newblock {\em Quantum}, 4:296, 2020.
\newblock \href {http://arxiv.org/abs/1902.10673} {\path{arXiv:1902.10673}},
  \href {https://doi.org/10.22331/q-2020-07-16-296}
  {\path{doi:10.22331/q-2020-07-16-296}}.

\bibitem{kivlichan2018quantum}
Ian~D Kivlichan, Jarrod McClean, Nathan Wiebe, Craig Gidney, Al{\'a}n
  Aspuru-Guzik, Garnet Kin-Lic Chan, and Ryan Babbush.
\newblock Quantum simulation of electronic structure with linear depth and
  connectivity.
\newblock {\em Physical Review Letters}, 120(11):110501, 2018.
\newblock \href {http://arxiv.org/abs/1711.04789} {\path{arXiv:1711.04789}},
  \href {https://doi.org/10.1103/PhysRevLett.120.110501}
  {\path{doi:10.1103/PhysRevLett.120.110501}}.

\bibitem{larkin2020evaluation}
Jason Larkin, Matías Jonsson, Daniel Justice, and Gian~Giacomo Guerreschi.
\newblock {Evaluation of QAOA based on the approximation ratio of individual
  samples}.
\newblock {\em arXiv e-prints}, 2020.
\newblock \href {http://arxiv.org/abs/2006.04831} {\path{arXiv:2006.04831}}.

\bibitem{li2020gibbs}
Li~Li, Minjie Fan, Marc Coram, Patrick Riley, and Stefan Leichenauer.
\newblock {Quantum optimization with a novel Gibbs objective function and
  ansatz architecture search}.
\newblock {\em Physical Review Research}, 2:023074, 2020.
\newblock \href {http://arxiv.org/abs/1909.07621} {\path{arXiv:1909.07621}},
  \href {https://doi.org/10.1103/physrevresearch.2.023074}
  {\path{doi:10.1103/physrevresearch.2.023074}}.

\bibitem{lykov2020tensor}
Danylo Lykov, Roman Schutski, Alexey Galda, Valerii Vinokur, and Yuri Alexeev.
\newblock Tensor network quantum simulator with step-dependent parallelization,
  2020.
\newblock URL: \url{https://arxiv.org/abs/2012.02430}, \href
  {https://doi.org/10.48550/ARXIV.2012.02430}
  {\path{doi:10.48550/ARXIV.2012.02430}}.

\bibitem{manurangsi2019}
Pasin Manurangsi.
\newblock {A Note on Max k-Vertex Cover: Faster FPT-AS, Smaller Approximate
  Kernel and Improved Approximation}.
\newblock In {\em 2nd Symposium on Simplicity in Algorithms, SOSA'19}, pages
  15:1--15:21, 2019.
\newblock \href {http://arxiv.org/abs/1810.03792} {\path{arXiv:1810.03792}},
  \href {https://doi.org/10.4230/OASIcs.SOSA.2019.15}
  {\path{doi:10.4230/OASIcs.SOSA.2019.15}}.

\bibitem{biamonte2018variational}
Mauro E.~S. Morales, Timur Tlyachev, and Jacob Biamonte.
\newblock {Variational learning of Grover's quantum search algorithm}.
\newblock {\em Physical Review A}, 98(6):062333, 2018.
\newblock \href {http://arxiv.org/abs/1805.09337} {\path{arXiv:1805.09337}},
  \href {https://doi.org/10.1103/PhysRevA.98.062333}
  {\path{doi:10.1103/PhysRevA.98.062333}}.

\bibitem{swap-networks}
Bryan O'Gorman, William~J. Huggins, Eleanor~G. Rieffel, and K.~Birgitta Whaley.
\newblock Generalized swap networks for near-term quantum computing.
\newblock {\em arXiv e-prints}, 2019.
\newblock \href{https://arxiv.org/abs/1905.05118}{arXiv:1905.05118}.
\newblock \href {https://doi.org/10.48550/ARXIV.1905.05118}
  {\path{doi:10.48550/ARXIV.1905.05118}}.

\bibitem{papadimitriou1991maxsnp}
Christos~H. Papadimitriou and Mihalis Yannakakis.
\newblock {Optimization, approximation, and complexity classes}.
\newblock {\em Journal of Computer and System Sciences}, 43(3):425--440, 1991.
\newblock \href {https://doi.org/10.1016/0022-0000(91)90023-X}
  {\path{doi:10.1016/0022-0000(91)90023-X}}.

\bibitem{petrank1994hardness}
Erez Petrank.
\newblock {The hardness of approximation: Gap location}.
\newblock {\em Computational Complexity}, 4(2):133--157, 1994.
\newblock \href {https://doi.org/10.1007/BF01202286}
  {\path{doi:10.1007/BF01202286}}.

\bibitem{verstraete2009quantum}
Frank Verstraete, J~Ignacio Cirac, and Jos{\'e}~I Latorre.
\newblock Quantum circuits for strongly correlated quantum systems.
\newblock {\em Physical Review A}, 79(3):032316, 2009.
\newblock \href {http://arxiv.org/abs/0804.1888} {\path{arXiv:0804.1888}},
  \href {https://doi.org/10.1103/PhysRevA.79.032316}
  {\path{doi:10.1103/PhysRevA.79.032316}}.

\bibitem{twos-complement}
John von Neumann.
\newblock {First Draft of a Report on the EDVAC}, 1945.

\bibitem{angles-basin-nasa}
Zhihui Wang, Nicholas~C. Rubin, Jason~M. Dominy, and Eleanor~G. Rieffel.
\newblock {$XY$ mixers: Analytical and numerical results for the quantum
  alternating operator ansatz}.
\newblock {\em Physical Review A}, 101(1):012320, 2020.
\newblock \href {https://doi.org/10.1103/PhysRevA.101.012320}
  {\path{doi:10.1103/PhysRevA.101.012320}}.

\bibitem{nasa2020XY}
Zhihui Wang, Nicholas~C. Rubin, Jason~M. Dominy, and Eleanor~G. Rieffel.
\newblock {$XY$ mixers: Analytical and numerical results for the quantum
  alternating operator ansatz}.
\newblock {\em Physical Review A}, 101(1):012320, 2020.
\newblock \href {http://arxiv.org/abs/1904.09314} {\path{arXiv:1904.09314}},
  \href {https://doi.org/10.1103/PhysRevA.101.012320}
  {\path{doi:10.1103/PhysRevA.101.012320}}.

\bibitem{wurtz2021pg1}
Jonathan Wurtz and Peter Love.
\newblock Maxcut quantum approximate optimization algorithm performance
  guarantees for $p>1$.
\newblock {\em Physical Review A}, 103:042612, 2021.
\newblock \href {http://arxiv.org/abs/2010.11209} {\path{arXiv:2010.11209}},
  \href {https://doi.org/10.1103/PhysRevA.103.042612}
  {\path{doi:10.1103/PhysRevA.103.042612}}.

\end{thebibliography}

\end{document}